\documentclass[10pt]{iopart}
\usepackage{iopams}
\usepackage{setstack}
\usepackage[dvips]{graphicx}
\usepackage{epsfig}

\usepackage{geometry}
\begin{document}

\title{Slow relaxation in microcanonical warming of a Ising lattice}

\author{E. Agliari$^{1,2,3}$, M. Casartelli$^{1,2}$ and A. Vezzani$^{1,4}$}
\address{$^1$ Dipartimento di Fisica, Universit\`a di Parma, Viale
G.P. Usberti n.7/A (Parco Area delle Scienze), 43100 - Parma - ITALY}
\address{$^2$ INFN, gruppo collegato di Parma, Viale G.P. Usberti
n.7/A (Parco Area delle Scienze), 43100 - Parma - ITALY}
\address{$^3$ Laboratoire de Physique Th\'eorique de la Mati\`ere Condens\'ee, Place Jussieu 4, 75252 - Paris - FRANCE}
\address{$^4$
Centro S3, CNR Istituto di Nanoscienze, via Campi 213A, 41125 - Modena - ITALY}

\begin{abstract}
We study the warming process of a semi-infinite cylindrical Ising lattice initially ordered and coupled at the boundary to a heat reservoir. The adoption of a proper microcanonical dynamics allows a detailed study of the time evolution of the system. As expected,  thermal propagation displays a diffusive character and the spatial correlations decay exponentially in the direction orthogonal to the heat flow. However, we show that the approach  to equilibrium  presents an unexpected slow behavior. In particular, when the thermostat is at infinite temperature, correlations decay to their asymptotic values by a power law. This can be rephrased in terms of a correlation length vanishing logarithmically with time. At finite temperature, the approach to equilibrium is also a power law, but the exponents depend on the temperature in a non-trivial way. This complex behavior could be explained in terms of two dynamical regimes characterizing finite and infinite temperatures, respectively.
When finite sizes are considered, we evidence the emergence of a much more rapid equilibration, and this confirms that the microcanonical dynamics can be successfully applied on finite structures. Indeed, the slowness exhibited by correlations in approaching the asymptotic values are expected to be related to the presence of an unsteady heat flow in an infinite system. 
\end{abstract}

\section{Introduction}
\label{intro}
The transient regime of an Ising system starting from equilibrium at a uniform temperature and then suddenly coupled to a thermostat at a different temperature,
is an issue involving both practical problems and  theoretical aspects of non equilibrium. 
For instance, as it is well known, cooling a system at a temperature smaller \cite{aging1,aging2,coarsening1,coarsening2} or equal \cite{godluck,calabrese} to the critical one, gives rise to interesting dynamical  out-equilibrium phenomena, where the transient regimes become indefinitely long. This approach, even for an artificial Monte Carlo dynamics, provides a deep insight in important physical problems, such as aging phenomena and coarsening of magnetic domains.

We remark that within the usual  canonical approach  used in these studies (e.g. Metropolis), the interaction with the reservoir yields a uniform thermalization, a fact avoiding heat transport. Therefore, in a canonical frame, the investigation of heat propagation during a out-equilibrium warming process is not even possible. This excludes  interesting issues related to transport: which are the relevant parameters of the warming process, the role of boundary conditions, the space-time scales, the evolution of correlations aligned or transversal to the heat flow, etc. \cite{libro1,malte,libro3,livi}. In this canonical context heat transport can be studied only at interface separating two systems at different fixed temperatures \cite{gonnella}.

Thus, in order to investigate heat transport phenomena in a transient regime, the appropriate dynamical approach should be the microcanonical one. This approach allows to couple the system to a thermostat and to investigate how the conserved energy flows within the system itself. However, so far, this approach has been largely neglected, possibly because of the need of a unique dynamical frame in a wide range of temperatures and/or of underlying geometries, not matched by standard microcanonical rules \cite{vic,pom,grant,saito1,creutz1,creutz2,acv2}. 

Now, a new efficient microcanonical dynamics has been recently introduced in \cite{ACV2009}, allowing for enhancements in the study of heat flow for a spin lattice coupled to thermostats at arbitrary fixed temperatures. Indeed, among the advantages of this dynamics (including e.g. the simple definition of local temperatures, the direct computability of conductivity, the possibility of applying to a great variety of discrete topologies and coupling patterns), there is also a property that the previous rules lacked, i.e. an efficient behavior in the whole range of temperatures from $0$ to $\infty$. 
We remark, that  previous results for such a dynamics on homogeneous or disordered systems, are limited to the {\sl stationary} regime, which is  established after a long-time interaction between the systems and the thermostats \cite{ACV2009,ACV2010,ACV2010E}. On the other hand, as already mentioned, here we focus on genuine non-stationary phenomena emerging when the system is suddenly put in contact with a thermostat at different temperature.

In particular, the system under study is an initially ordered rectangular lattice, where at time $0$ one side is coupled to a reservoir at fixed temperature $T$, and consequently a heat diffusion takes place in the direction orthogonal to the hot side. This rectangle, in order to have a permanent state of transience, should be thought of as infinite in the diffusing direction. In practice, we shall take this size $L$ greater than the grid thermalized by the expanding heat front during a long observation time. Periodic boundary conditions are assumed in the transverse direction, so that the geometry reduces to a semi-infinite cylinder.
 
Some basic concepts of out-equilibrium physics, which have been developed for instance in the study of coarsening process, will be very useful also in our approach. In particular, dynamical scaling is a general tool that on one hand gives a simple description of the system evolution, and on the other it provides a theoretical basis for the introduction of universal concepts, such as correlation length and dynamical exponents. In this perspective, since for our dynamics a diffusive behavior has been proven \cite{ACV2009,ACV2010,ACV2010E}, a generic observable $\mathcal{O}$ is expected to depend on the time  $t$ and on the distance $x$ from the reservoir according to the dynamical scaling: 
$$
\mathcal{O}(x,t) \sim f(\tilde t) ~, \quad ~{\tilde t} = t/x^2 ~.
$$
Notice that the rescaling procedure allows to introduce a stationary approach even in the transient problem; indeed, data obtained at different times can be compared to each other by a simple redefinition of distances:
$$
{\tilde x}=x/t^{1/2} \propto x/\ell(t) ~, 
$$
where $\ell(t) $ is a characteristic diffusive length.

The most interesting feature we will evidence is that, in the presence of non-stationary heat propagation, the approach to equilibrium may be very slow, even when heating a system at high temperature.

In this context an interesting observable is the correlation $\langle \sigma_{x,y}(t)~ \sigma_{x,y+n}(t) \rangle  $, measured at time $t$ between the value of spins at sites $y$ and $y+n$ of a line orthogonal to the heat flow at distance $x$ from the hot side. For $n=1$ this correlation corresponds to the magnetic energy. Since we are interested in the relaxation towards the equilibrium, we shall consider the quantity ${\cal C}_n(x,t) \equiv \langle \sigma_{x,y}(t) ~\sigma_{x,y+n}(t) \rangle - \langle \sigma_{x,y} \sigma_{x,y+n} \rangle$, where $\langle \sigma_{x,y} \sigma_{x,y+n} \rangle$ is the correlation function of a system at equilibrium at the temperature of the heat reservoir. Thus, ${\cal C}_n(x,t)$ tends to 0 in the limit $t \to \infty$; moreover, according to the scaling ansatz ${\cal C}_n(t,x)$ is a function $F_n(\tilde{t})$. 

By numerical simulations, first we check the scaling hypothesis, then we study the behavior of the scaling functions $F_n({\tilde t})$. In particular, after extensive Monte Carlo simulations, we obtain $F_n \sim \tilde t^{-\alpha(n,T)} $, where $\alpha(n,T) = \alpha_1(T) n + \alpha_0(T) $. In this context two simple behaviors emerge, namely $\alpha(n,T) \sim n/2$ and $\alpha(n,T) \sim 1/2$, for large and  small temperatures respectively. We will show that a dynamics characterized by $\alpha(n,T) = n/2$ provides a reasonable description of the system evolution at infinite temperature. On the other hand, $\alpha(n,T) = 1/2$  is expected to be a good behavior for finite temperature systems. In this perspective the complex dependence on $T$ displayed by the exponents $\alpha(n,T)$ emerges as due to a dynamical crossover between the dynamics pertaining to finite and infinite temperature respectively.

We notice that the diffusive behavior characterizing the dynamical scaling, as well as the exponential decay of the correlations as a function of $n$, are expected features of the model; indeed, the correlation decay is typically exponential in the Ising model as soon as the magnetic domains are regular, i.e. outside the critical regime. On the other hand, the way the correlations tend to their equilibrium values, i.e. by a power law in $\tilde t$, is not so obvious. It seems that the system persists in a transient regime even for times much larger than  the characteristic timescale of diffusion, which is proportional to $x^2$. In some way, even at fixed distance from the thermostat there is not a characteristic time for which the  equilibrium could be considered as reached. 

We will finally show that such a slow relaxation is a typical feature of open infinite systems where non-steady heat transport is present for all times. Indeed, in a finite system, a rapid decay to the equilibrium value takes place as soon as the diffusion characteristic length $ \ell(t) $ is of the order of magnitude of the system size $L$. For this reason the dynamics considered here can be adopted for the study of the stationary properties of finite systems, as it was done in \cite{ACV2009,ACV2010,ACV2010E}.

The paper is organized as follows: In Sec.~\ref{microdyn} we briefly review the microcanonical dynamics exploited to simulate the evolution of the system, then, in Sec.~\ref{sec:numerics}  we present the numerical results focusing on diffusion features, on correlations and on finite-size effects; finally, Sec.~\ref{sec:concl} is left for conclusions. 

\section{Microcanonical Dynamics and Model}
\label{microdyn}
In order to be self-contained, we briefly recall the dynamics introduced in \cite{ACV2009},
whose main novelty consists in assigning to the \emph{links}, beyond the usual magnetic energy $E^m_{ij}$, a sort of ``kinetic'' term $E^k_{ij}$ which is a non-negative definite quantity able to compensate the variations of magnetic energy due to the spin flips involving the adjacent nodes $i$ and  $j$. This makes an essential difference with respect to the Creutz microcanonical procedure \cite{creutz1,creutz2}, where a \emph{bounded} amount of kinetic energy is assigned to each \emph{node}: in that case, indeed, both the energy boundedness and the connectivity of the nodes entail many limitations, ranging from  geometrical or topological constraints to  non-ergodicity at low energy density. 

Such limitations, as shown in \cite{ACV2009}, are efficiently solved by the new dynamics. An elementary move for it, starting from a random distribution of link energies, consists in the following:
 \begin{enumerate}
 \item select randomly one of the  four spin configurations for a randomly extracted link  $\{i,j\}$, and evaluate the magnetic energy variation $\Delta E^m_{ij}$ due to this choice. If the coupling $J_{ij} \equiv  1$, as in the present case,  $\Delta E^m_{ij}$ is an integer whose value depends on the connectivity of the nodes $i$ and $j$;

 \item if $\Delta E^m _{ij}\leq 0$, accept the choice and increase the link kinetic energy $E^k_{ij}$ of $\Delta E^m _{ij}$; %
if $\Delta E^m _{ij}> 0$, accept the choice and decrease $E^k_{ij}$  of $\Delta E^m _{ij}$ only if the link energy remains non negative, otherwise reject the choice.
  \end{enumerate}
The unit time step consists of $N$ moves, where $N$ is the total number of links.
Since a link is defined only by the adjacent nodes, this rule applies on arbitrary non-oriented
connected graphs, providing a great generality to the dynamics.

We remark that at low energies the dynamics is rather time-consuming as for most of the extracted links the move is typically rejected; yet, differently from the Creutz procedure, here the dynamics works at any energy.

 In \cite{ACV2009,ACV2010}, where a cylindrical lattice was considered, many points have been numerically or theoretically supported, for both homogeneous and disordered cases. For instance, starting from the fact that magnetic and kinetic energies are non-correlated observables, that the Boltzmann distribution is recovered at equilibrium and that the system results to be ergodic at all finite temperatures, it is possible to define the temperature $T_{ij}$ by the averaged  kinetic energy $ \langle E_{ij}^k \rangle$: this is a link observable, and therefore a local quantity working as a temperature also in stationary states far from equilibrium, i.e. states forced by thermostats at different temperatures.

\section{Results}
\label{sec:numerics}

In order to study the transient regime, we deal here with the same simple cylindrical geometry considered  in \cite{ACV2009,ACV2010}. It consists of a rectangular lattice of $L_X \times L_Y$ sites, periodical in the $Y$ direction and open in the $X$ direction: the first and last columns, discrete circles of $L_Y$ sites, are therefore open borders, in contact with heat reservoirs constituted by supplementary columns (two for each side are enough). These reservoirs are kept at constant temperatures $T_1$ and $T_2$ by using the Metropolis dynamics at equilibrium with appropriate temperatures. In most cases, for our present exigences, only the thermostat at the left border is fixed, while the right border is open and virtually placed at infinity (i.e. very large $L_X$).
In the following we report and discuss the results obtained by means of Monte Carlo (MC) simulations performed on cylinders %of size $L_X \times L_Y$, 
initially set at zero temperature, that is displaying an ordered magnetic configuration and zero kinetic energy on links.  
A given site $i$ will be denoted by the couple $\{x,y\}$, where the integers $x$ and $y$ represent the distance from the reservoir and the position along the column respectively. 
Moreover, in order to focus on the main features of the transient, presently we limit to the homogeneous case, i.e. the coupling constants are set equal to $1$ for any couple of neighboring sites.

Time is measured in MC steps, each corresponding to a total extraction of $L_X \times L_Y$ links, with consequent operations $1.$ and $2.$, as described in the previous section. In this way we get temporal series for the main observables, such as the average magnetization $M(x,t)$ and the average energy $E(x,t)$, for the $x$-th column at time $t$, being
\begin{eqnarray}
M(x,t) = \frac{1}{L_Y} \sum_{y=1}^{L_Y} \langle \sigma_{x,y}(t)\rangle,\\
E(x,t) = - \frac{1}{L_Y} \sum_{y=1}^{L_Y} \langle \sigma_{x,y}(t)\sigma_{x,y+1}(t) \rangle,
\end{eqnarray}
where $\sigma_{x,y}$ is the value of the spin on site $(x,y)$ at time $t$ and $\langle . \rangle$ is the average over different thermal realizations (typically in our simulation we average over $10^4$ thermal histories). 
Thus, as the simulation is onset, heat starts to propagate within the cylinder: the absolute average magnetization decreases and the average energies of columns close to the thermostat increase.

The first part of simulations is devoted to the analysis of heat propagation, and the longitudinal size is taken rather large and equal to $L_X=300$, so that heat can flow for long times without feeling the influence of the second border. In the following part we focus on finite-size effects and we consider relatively small longitudinal sizes, that is $L_X=20$. In any case, $L_Y$ is kept fixed and equal to $300$; this large value is useful to avoid finite size effects and improves the statistics of the measures.

\subsection{Diffusion features}

\begin{figure}
\begin{center}
\resizebox{0.6\columnwidth}{!}{
\includegraphics{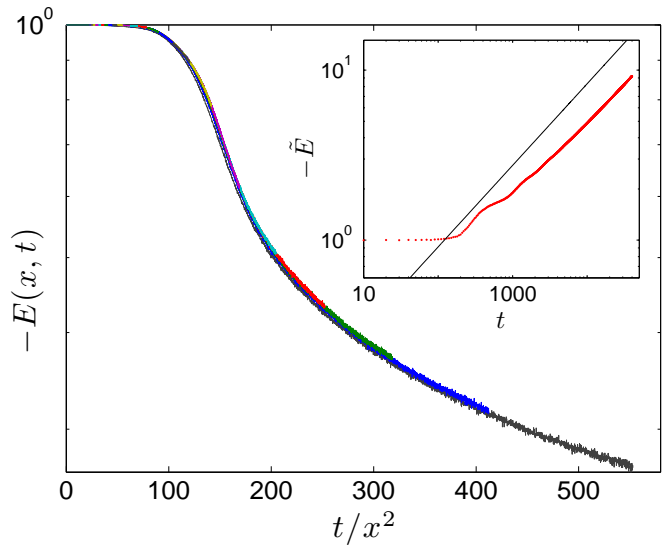}}
\caption{
(Color on line) Main figure: Each curve, depicted in different color, represents the magnitude of the mean energy $- {E}(x,t)$,  as a function of the rescaled time $\tilde{t}$. Different columns of the system $x$  have been taken into account. Inset: the energy $\tilde{E}$ (red symbols) as a function of time $t$, compared with the power law $\sim t^{1/2}$ (black line).  Numerical data can be fitted by a power-law with exponent $0.48$; the small difference with respect to the expected value $0.5$ can be ascribed to the existence of a long pre-asymptotic regime. All data 
represented in this plot refer to a system of size $L_Y=300$ initially ordered and subject to a thermostat at $T=10000$.
}
\end{center}
\label{fig:diffusione} 
\end{figure}

In the presence of a static heat flux, the system behaves diffusively, as already noticed in \cite{ACV2009,ACV2010}. The present simulations confirm this picture even in a transient regime. In particular, relevant physical quantities such as the total energy along the $Y$ direction 
%\begin{equation}
%\nonumber
$$
\tilde{E}(t) = \sum_{x=1}^{L_X} E(x,t),
$$
%\nonumber
%\end{equation}
grows  as $t^{1/2}$, that is the characteristic law of diffusive processes; this asymptotic regime is reached only at very large times due to the extensive  routines of the dynamics (see inset of Fig. \ref{fig:diffusione}).
Moreover, Fig.~\ref{fig:diffusione}  evidences that focusing on the direction of heat propagation, the system can be studied within the framework of dynamical scaling which is typical of out-equilibrium systems. More precisely, the quantities (e.g. the local energy) depending on both time $t$ and distance $x$ can be plotted in terms of a single rescaled variable $\tilde x= x/\ell(t)$, where $\ell(t)\sim t^{1/2}$ is the characteristic length of the process, so that data obtained at different spaces and times can be plotted together on a single curve.
The fact that in our system $\ell(t)$ grows as $t^{1/2}$ confirms the diffusive behavior of the model.
Of course, an equivalent scaling (which has been used in our plot) consists in introducing a rescaled time  $\tilde t = t/x^2$. 
Moreover, we remark that the scaling approach is very general and does not depend on the observable we have considered. Indeed we have checked that different observables, such as the magnetization and the correlation functions, display an analogous diffusive dynamical scaling. 

Finally, it is worth deepening the relation between the dynamical
scaling introduced above and the generalized Fourier equation used in \cite{ACV2009} for the description of the model
\begin{equation} \label{eq:diff}
\frac{\partial E}{\partial t} = \frac{\partial }{\partial x} \left( D(E)
\frac{\partial E}{\partial x}  \right),
\end{equation}
where $D(E)$ is a diffusivity factor which, in quasi-equilibrium states, is just the ratio between the conductivity and the specific heat, so that the standard Fourier equation is recovered. We recall that the previous equation is able to describe the
heat propagation in very general conditions, independently of the existence of a
local temperature \cite{CVM}. Now, it is straightforward to verify that posing
$E(x,t)= E(\tilde{x})$ leads to
\begin{equation} \label{eq:diff2}
- \frac{\tilde{x}}{2}  \frac{\partial E}{\partial \tilde{x}} = \frac{\partial
}{\partial \tilde{x}} \left( D(E) \frac{\partial E}{\partial \tilde{x}}  \right),
\end{equation}
i.e. the equation can be rewritten in terms of $\tilde x$ only.
Moreover, assuming that $D(E)$ is a sufficiently smooth function writable as
$D(E)=D_0+D_1 E$, from Eq.~\ref{eq:diff2} we can derive that the energy can be
written as $E(\tilde{x}) = E_0 + E_1 \tilde{x}$, namely $E(t,x) - E_0 \sim x/
\sqrt{t}$ (at least at long times). As we will see in the following subsection, this estimate is consistent with our numerical results.

\subsection{Correlations}

The transient regime presents interesting properties not only along the direction of heat flow but also along the orthogonal lines. In this framework, the simplest observable to consider is the two-point correlation function between spins of the same column at distance $n$, i.e.:
\begin{equation}
C_n(x,t) = \frac{1}{L_Y} \sum_{y=1}^{L_Y} \langle \sigma_{x,y}(t)\sigma_{x,y+n}(t) \rangle ~.
\end{equation}
Clearly, $C_1(x,t)$ is simply the negative of the energy $E(x,t)$ of the $x$-th column, i.e. $C_1(x,t) = -E(x,t)$. Since we are interested in the long-time asymptotic behavior, we consider the quantity
\begin{equation}
{\cal C}_n(x,t) =  C_n(x,t) - C_n^{\infty},
\end{equation}
where $C_n^{\infty}=\langle \sigma_{x,y} \sigma_{x,y+n} \rangle=\langle \sigma_{x,y} \sigma_{x+n,y} \rangle$  is the equilibrium two-point correlation function between spins at distance $n$ in a 2-dimensional system  evaluated at the same temperature of the thermostat. Therefore, if the system approaches to the equilibrium, we have $\lim_{t\to \infty} {\cal C}_n(x,t)=0$, as it has been indeed verified in all our numerical simulations.

 In the Ising model  outside the critical regime, when simply connected domains are prevalent, the correlation functions decay exponentially with the distance $n$, i.e. as $\exp(-n/\lambda)$, where $\lambda$ is defined as the correlation length. At equilibrium, $\lambda$ depends on the temperature, while, in the coarsening processes, it grows as $t^{1/2}$, due to the diffusion of the magnetic domains \cite{aging1,coarsening1,coarsening2}.

\begin{figure}
\begin{center}
\resizebox{0.6\columnwidth}{!}{\includegraphics{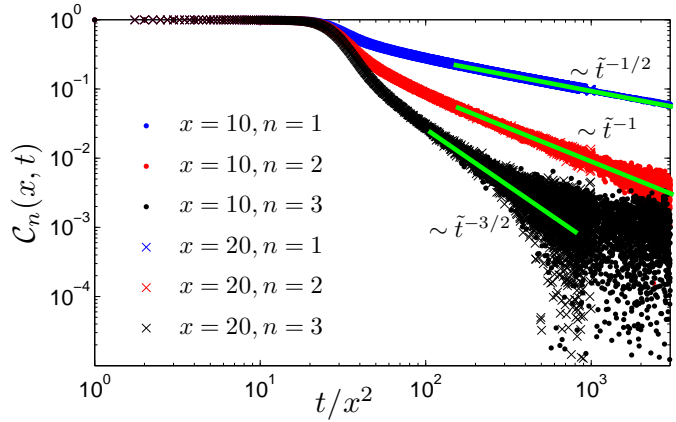}}
\caption{(Color on line) Log-log scale plot of the correlation function $\mathcal{C}_n$, for $n=1$ (blue), $n=2$ (red) and $n=3$ (black), as a function of the rescaled time $\tilde{t} = t/x^2$. Measures have been performed on two columns, at distance $10$ ($\bullet$) and $20$ ($\times$) from the thermostat, respectively; due to time rescaling, data pertaining to different columns are collapsed. The green curves represent power-laws functions $\sim t^{-n/2}$ and provide a good fit for experimental data, over a proper time range.
All data represented in this plot refer to a system of size $L_Y=300 $ initially ordered and subject to a thermostat at temperature $T=10000$.
}
\end{center}
\label{fig:tri}
\end{figure}

Fig. \ref{fig:tri} shows the behavior of ${\cal C}_n(x,t)$ when the thermostat is at very large temperature, i.e. infinite for all practical purposes. In this case the equilibrium correlation function vanishes. The dynamical scaling picture is verified, since data evaluated at different columns (e.g. $x=10$ and $x=20$) collapse by using the rescaled variable 
$\tilde t =t/x^2$, i.e. ${\cal C}_n(x,t)=F_n(\tilde t)$. Moreover,  for large enough $\tilde t$, simulations evidence that 
\begin{equation}
\label{scal_inf}
F_n(\tilde t)\sim \tilde t^{-n/2}. 
\end{equation}
Therefore, the correlation function presents an exponential decay with the distance $n$ and this is peculiar of the Ising model outside the critical regime.

On the other hand, an unexpected and interesting feature is the slow decay of $F_n(\tilde t)$ as a function of $\tilde t$, consistently with the predictions drawn in the previous section for $n=1$.
While in the standard MC approach a slow equilibration is typical of {\sl freezing} only (below the critical temperature), here we can appreciate a similar behavior also during the warming phase. In fact, a rapid equilibration would involve an exponential decay in $\tilde t$. In particular, the heating process can not be considered concluded even for those sites whose distance from the thermostat is much smaller than $\ell(t)$, i.e. the characteristic length of heat diffusion. From a different point of view, the very slow approach to equilibrium can be described in terms of correlation length $\lambda$ which, in our model, turns out to tend to zero logarithmically in $\tilde t$; indeed the correlation function can be written as:
\begin{equation}
{\cal C}_n(x,t) = F_n(\tilde t)\sim \tilde t^{-n/2} = \exp \left[ {-\frac{n \log(\tilde t)}{2}} \right] = \exp(-n/\lambda). 
\label{logdecay}
\end{equation}
 We will show that the unexpected slow relaxation at high temperature emerges in open systems, while in the next section we verify that in small systems, where the  boundary effects can not be discarded and heat can not flow freely, the thermalization process is exponentially fast at large enough times.

The scaling function in Eq.~\ref{logdecay} presents a very simple structure. Let us define $g_n(\tilde x) \equiv F_n(\tilde t)$ and expand $g_n(\tilde x)$ for small $\tilde x$ (i.e. large times); we get
\begin{equation}
g_n(\tilde x) = \sum_{k=0}^{\infty} A_{k,n} \tilde x^k.
\end{equation}
Then, formula (\ref{scal_inf}) can be recast as $A_{k,n}=0$ if $k<n$, giving  a description,  at infinite temperature, of the asymptotic behavior of the model in terms of the analytic properties of $g_n(\tilde x)$.
Now, when the border is thermalized at a finite temperature, the main features characterizing the heat transport are preserved: as evidenced in Fig.~$3$, scaling still holds with a definite diffusive  length, the correlation functions decay exponentially and the approach to equilibrium is  slow since the scaling functions vanish as a power law. However, here the analytic form of $F_n(\tilde t)$ is not so simple; in fact we have $F_n \sim \tilde t^{-\alpha(n,T)} $, and fitting procedures suggest that 
\begin{equation}
\alpha(n,T) = \alpha_1(T) n + \alpha_0(T)
\end{equation}
as shown in the inset of Fig.~$3$.

\begin{figure}
\begin{center}
\resizebox{0.6\columnwidth}{!}{
\includegraphics{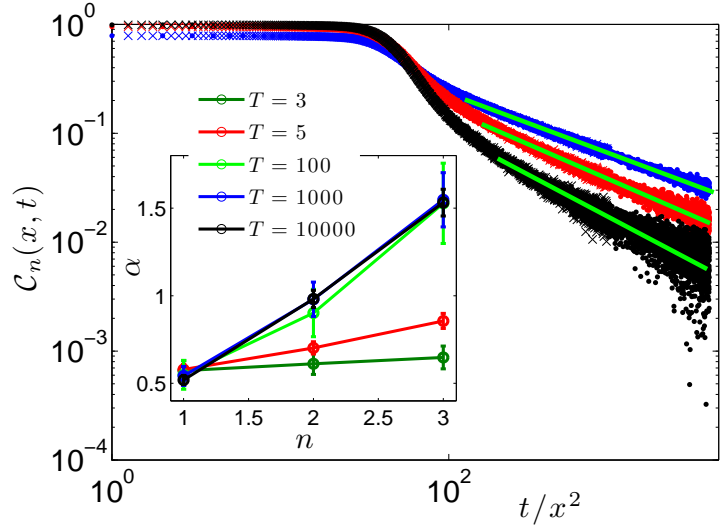}}
\caption{(Color on line) Main figure: log-log scale plot of the correlation function $\mathcal{C}_n$, for $n=1$ (blue), $n=2$ (red) and $n=3$ (black), as a function of the rescaled time $\tilde{t}$, for a system of size $L_Y=300$ initially ordered and subject to a thermostat at temperature $T=5$.
Measures have been performed on two columns, at distance $10$ ($\bullet$) and $20$ ($\times$) from the thermostat, respectively; due to time rescaling, data pertaining to different columns are collapsed. The green curves represent power-laws functions $\sim t^{- \alpha(n,T)}$ and provide a good fit for experimental data, over a proper time range.
Inset: Exponent $\alpha(n,T)$ obtained by fitting data from numerical simulations, as a function of $n$. Different temperatures for the thermostat are considered and compared, as explained by the legend.
}
\end{center}
\label{fig:collassat5}
\end{figure}

\subsection{Dynamical regimes}

The  exponents characterizing the slow decays appear to be temperature dependent with a non-trivial behavior. However,  Fig.~$3$ evidences that  for high (practically infinite) temperature, $\alpha(n,T) = n/2$, while, for low enough temperatures, $\alpha(n,T) \sim 1/2$ independently of $n$. This suggests that two simple (asymptotic) behaviors are present in the model: for infinite temperature the scaling function should be given by
\begin{equation}
\label{Tinf} 
F_n(\tilde t)\sim \tilde t^{-n/2}= \exp \left[ - \frac{ n\,   \log(\tilde t) }{2} \right],
\end{equation}
hence, at infinite temperature, the dynamics is characterized by indefinitely shortening correlation length $\lambda(\tilde{t})=2/\log(\tilde t)$. This is clearly consistent with the fact that at infinite temperature there is not an equilibrium characteristic length, since the correlation function between any couple of sites is vanishing.
On the other hand, at small temperature,  $F_n(\tilde t)$ decays as $\tilde t^{-1/2}$. However, since the exponential decay of $C_n(t)$ as a function of the distance $n$ is an expected feature, we have  
\begin{equation}
\label{Tfin} 
F_n(\tilde t)\sim \tilde t^{-1/2} \exp(- n/\lambda')
\end{equation}
where a characteristic length $\lambda'$ naturally emerges. We remark that at finite temperature the equilibrium correlation functions decay with a characteristic length $\lambda_{eq}(T)$. Since a similar behavior is expected to be reached asymptotically also out of equilibrium, we expect that, eventually, $\lambda'\sim \lambda_{eq}(T)$. Therefore the above regimes (\ref{Tinf},\ref{Tfin}) provide a reasonable description for infinite and finite temperature respectively. 

In this perspective, the complex behavior characterizing intermediate temperatures would result from an interplay between these regimes. In particular, when the thermostat is at a finite temperature $T$, $F_n(\tilde{t})$ evolves according to Eq. (\ref{Tinf}) until the correlation length $\lambda(\tilde{t})$ becomes comparable to $\lambda'\sim \lambda_{eq}(T)$; afterwards an evolution described by (\ref{Tfin}) takes place. 
%In this framework a system at a temperature much smaller than $T$ evolves as if the thermostat were at infinite temperature and only when $
%\lambda(t) \sim \lambda_{eq}(T)$, effects due to small temperature difference can take place as described by (\ref{Tfin}). 
At very large temperatures $\lambda'\sim \lambda_{eq}(T)$ is almost zero and, in the considered time regimes, $\lambda(\tilde{t})\gg \lambda'$, therefore only the infinite temperature evolution (\ref{Tinf}) is observed. At low temperatures the correlation $\lambda'$ is so large that $\lambda(\tilde{t})$ becomes quickly comparable to $\lambda'$ and $F_n({\tilde{t}})$ evolves according to (\ref{Tfin}). Finally, the temperature dependent exponents $\alpha(n,T)$ observed at intermediate temperatures should be the result of the crossover between the two regimes which is ongoing in the considered time window.

\subsection{Finite-size effects}

Let us now discuss the finite-size effects in the processes of heat transfer from the thermostat to the system. Finite-size effects can clarify some features outlined in the previous analysis, in particular the mutual influence  between the transport regime and the way the asymptotic values are approached.

Fig. $4$ shows the behavior of $\mathcal{C}_1(x,t)$ as a function of time for the column $x=10$ and a size $L_X=20$. Differently from the behavior depicted in Fig.~\ref{fig:diffusione}, here the dynamics speeds up, approaching quickly the equilibrium value, for large enough times, i.e. when $\ell(t)$ is comparable to the lattice size. Therefore, the presence of very long transients described by a power law seems to be typical of infinite systems, while finite systems thermalize in a rapid  (possibly exponential) way.  This differences ensure that the dynamics may be used for the study of the stationary properties of finite systems as it has been done in \cite{ACV2009,ACV2010,ACV2010E}. Moreover, Fig.~$4$ evidences that the characteristic time for equilibration increases with the temperature becoming very large at $T=\infty$.

\begin{figure}
\begin{center}
\label{fig:finite} 
\resizebox{0.6\columnwidth}{!}{
\includegraphics{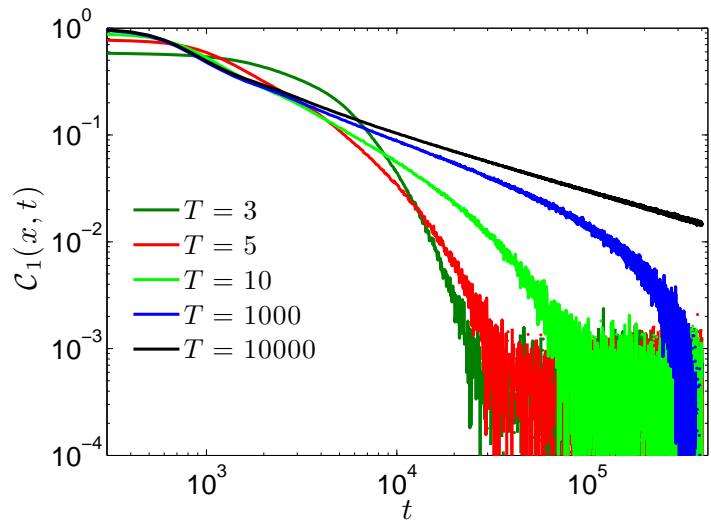}}
\caption{(Color on line) Correlation function $\mathcal{C}_1$ as a function of time $t$, for a system of size $L_Y=300$, $L_X=20$ and $x=10$, subject to a thermostat set at different temperature, each temperature being represented by a different color as explained by the legend.}
\end{center}
\end{figure}

\section{Conclusions }
\label{sec:concl}

The dynamical insight on the warming process in the Ising model we have studied has two main aspects: first, a direct confirmation of such expected features as the diffusive behavior in the heat propagation, or the exponential decay of correlations in the transversal direction; second, a clear indication about the role of heat flux in keeping the time behavior of correlations well distinct from their time-asymptotic ``equilibrium'' value. Indeed, in the indefinite propagation taking place when the longitudinal size is much longer than the observation window, the approach to the limit obeys a power law in time. Remarkably, a slow relaxation seems to be present for all temperatures. In fact,  one could in principle conjecture the existence of a special temperature (possibly related to the critical one) separating a slow and a fast dynamical regime; yet the persistent slowness of the dynamics proves the non-existence of such a dynamical threshold. In particular, the explanation of the non-trivial behavior of the temperature dependent exponents, in terms of finite and infinite temperature regimes, confirms the fact that other intermediate thresholds do not exist.

\section*{Acknowledgments}
This work has been partially supported by the MIUR Project PRIN 2008 ``Nonlinearity and disorder in classical and quantum processes'', 
and by the FIRB project RBFR08EKEV. EA is grateful to the Italian Foundation ``Angelo della Riccia''  for financial support.

\end{document}